\documentclass[journal]{IEEEtran}

\usepackage{float}
\usepackage{amsmath}
\usepackage{mathrsfs}
\usepackage{booktabs}
\usepackage{threeparttable}
\usepackage{algorithm, algorithmic}
\usepackage{multirow}
\usepackage{longtable}
\usepackage{graphics}
\usepackage{graphicx}
\usepackage{diagbox}
\usepackage[usestackEOL]{stackengine}
\usepackage{amssymb}

\ifCLASSINFOpdf
\else
\fi
\begin{document}
%
\title{Computing the Number of Equivalent Classes on $\mathcal{R}(s,n)/\mathcal{R}(k,n)$}
%
%
%


\author{Xiao~Zeng~\IEEEmembership{} and~Guowu~Yang ~\IEEEmembership{}
	\IEEEcompsocitemizethanks{\IEEEcompsocthanksitem X. Zeng and G. Yang are with the Big Data Research Center, University of Electronic Science and Technology of China, Chengdu 611731, China.\protect\\
		E-mail: xiaozeng@std.uestc.edu.cn, guowu@uestc.edu.cn. 
		(Corresponding author: Guowu Yang.)}
}

\maketitle

\begin{abstract}
Affine equivalent classes of Boolean functions have many applications in modern cryptography and circuit design. Previous publications have shown that affine equivalence on the entire space of Boolean functions can be computed up to 10 variables, but not on the quotient Boolean function space modulo functions of different degrees. Computing the number of equivalent classes of cosets of Reed-Muller code $\mathcal{R}(1,n)$ is equivalent to classifying Boolean functions modulo linear functions, which can be computed only when $n\leq 7$. Based on the linear representation of the affine group $\mathcal{AGL}(n,2)$ on $\mathcal{R}(s,n)/\mathcal{R}(k,n)$, we obtain a useful counting formula to compute the number of equivalent classes. Instead of computing the conjugate classes and representatives directly in $\mathcal{AGL}(n,2)$, we reduce the computation complexity by introducing an isomorphic permutation group $P_n$ and performing the computation in $P_n$. With the proposed algorithm, the number of equivalent classes of cosets of $R(1,n)$ can be computed up to 10 variables. Furthermore, the number of equivalent classes on $\mathcal{R}(s,n)/\mathcal{R}(k,n)$ can also be computed when $-1\leq k< s\leq n\leq 10$, which is a major improvement and advancement comparing to previous methods.
\end{abstract}

\begin{IEEEkeywords}
Boolean functions, affine equivalence, algebraic degree, group representation
\end{IEEEkeywords}

%
\IEEEpeerreviewmaketitle

\section{Introduction}
%
%
%
%
\IEEEPARstart
{A}{ffine} equivalence of Boolean functions have many applications in computer science and modern cryptography. The theory of affine equivalence is a useful tool for the design of logic networks that contain exclusive-OR modules among the set of primitive building blocks \cite{tc3}. It also has critical applications in cryptography, such as the classifications of cosets of Reed-Muller codes \cite{tc1}, \cite{tc2}, \cite{addtc1}, \cite{b5}, \cite{xdh}, \cite{b6}, \cite{b11}, and S-boxes \cite{b7}, \cite{sbox1}, \cite{b8}. 
\par An $n$-variable Boolean function is a mapping from $n$ binary bits to one binary bit, denoted as $f(x):\mathcal{F}_2^n \rightarrow \mathcal{F}_2$.  Two Boolean functions $f$ and $h$ are affine equivalent if there exists an invertible matrix $A$ and an n-dimensional vector $b$ over the Boolean field $\mathcal{F}_2$ such that $h(x)=f(Ax\oplus b)$, where $x$ is the input $(x_1,x_2,\cdots,x_n)^T$. The addition and multiplication are calculated in $\mathcal{F}_2$. Group techniques have been applied to analyze the affine equivalence of Boolean functions \cite{zhangyan}, \cite{siam}. Yan Zhang computed the affine equivalent classification of Boolean functions up to 10 variables by group isomorphism \cite{zhangyan}. In fact, the definition of equivalence can be extended by considering Boolean functions with certain degrees instead of the whole Boolean functions space. Two Boolean functions $f,h$ whose degrees are between $k$ and $s$ are equivalent if there exists an invertible matrix $A$ and an $n$-dimensional vector $b$ over $\mathcal{F}_2$ such that $h(x)=f(Ax\oplus b)$ mod $\mathcal{R}(k,n)$. This kind of equivalence is related to the classification of the cosets of Reed-Muller codes. 
It is worth mentioning that matrix techniques can be useful to analyzing some problems concerned with Boolean functions. Some Boolean control networks problems can be converted into the solvable set of Boolean matrix equations equivalently \cite{a1}, \cite{a2}.
\par Affine equivalence plays an important role in cryptography, since many cryptographic properties of Boolean functions remain the same after affine transformations \cite{b10}, \cite{add luccio}.  Algebraic degree, as a key cryptographic property of Boolean functions, has many applications in the area \cite{zwg}, \cite{zwg2}. Affine equivalence can be used to analyze nonlinearity and the algebraic degree of Boolean functions \cite{it1}, \cite{zfr}, \cite{zfr2}. It has also been successfully used in the classification of Reed-Muller codes, due to the fact that when two Boolean functions $f$ and $h$ are affine equivalent, they are also in the same cosets of Reed-Muller codes. Coset is a fundamental object studied in linear coding theory \cite{b16}, due to its strong connections with syndromes. The classification of cosets of the first order Reed-Muller code is equivalent to classify Boolean functions modulo linear functions under the definition. To be specific, two Boolean functions $f$ and $h$ are affine equivalent, if there exists an invertible matrix $A$ over $F_2$, $b,c\in \mathcal{F}^n_2, d\in \mathcal{F}_2$ such that $h(x)=f(Ax\oplus b)\oplus c^Tx\oplus d$.

\par Affine equivalence of rotation symmetric quadratic or cubic Boolean functions has been studied in \cite{add symmetry trotaion}, \cite{add1}, \cite{add2}, \cite{add3}. However, the affine equivalent classification of  Boolean functions with different degrees has been studied only by a few scholars. Using group isomorphism and Burnside's lemma, Yan Zhang \cite{zhangyan} computed affine classification up to 10 variables. This method cannot deal with the extended equivalent classification related to the classification of Reed-Muller codes, however. In fact, there are few studies that compute the whole classification of the cosets of Reed-Muller codes. In 1972, Berlekamp and Welch \cite{b11} successfully classified all 5-variable Boolean functions into 48 equivalent classes. In 1991, Maiorana \cite{b4} classified all 6-variable Boolean functions into 150357 equivalent classes by decomposing Boolean functions into $5$-variable functions and then applying the method for $5$-variable Boolean functions. In \cite{b6}, An Braeken computed the classification of the affine equivalent classes of cosets of the first order Reed-Muller code with respect to cryptographic properties such as correlation immunity, resiliency and propagation characteristics. They also determined the classification of $R(3, 7)/R(1, 7)$. In \cite{xdh}, the author proposed a method to compute the number of equivalent classes on $\mathcal{R}(s,n)/\mathcal{R}(k,n)$, and the number of variables are limited to 7. Based on the theory of $\mathcal{AGL}(n,2)$ acting on $\mathcal{R}(s,n)/\mathcal{R}(k,n)$ \cite{b16}, \cite{xdh}, we can compute the number of equivalent classes up to 10 variables using our own techniques in this paper.


The main objective of this paper is to computing the number of equivalent classes on $\mathcal{R}(s,n)/\mathcal{R}(k,n)$. By introducing the algebraic normal form of Boolean functions, we can represent a Boolean function as a unique coefficient vector, which means that there is a bijection between Boolean functions and the coefficient vector space. Then, the group representations of the affine group $\mathcal{AGL}(n,2)$ on $\mathcal{R}(s,n)/\mathcal{R}(k,n)$ are naturally established, which allows us to analyze the problem from the perspective of matrices. Using group representation of the affine group and Burnside's lemma, an useful counting formula is obtained to computing the number of equivalent classes on $\mathcal{R}(s,n)/\mathcal{R}(k,n)$. But computing the conjugate classes of the affine group $\mathcal{AGL}(n,2)$ can be very difficult and source consuming when the number of variables $n$ is large. We reduce the complexity by transforming the computation of conjugate classes of $\mathcal{AGL}(n,2)$ into the computation in a permutation group $P_n$, which was constructed in \cite{zhangyan} and is isomorphic to $\mathcal{AGL}(n,2)$. By computing the conjugate classes in $P_n$ and then transforming them into the conjugate classes in $\mathcal{AGL}(n,2)$, we can compute the affine equivalent classes up to 10 variables, which is a major advancement over previous methods.

The rest of this paper is organized as follows. Section 2 introduces some basic definitions and theories related to this paper. In Section 3, the counting formula of number of orbits of $\mathcal{AGL}(n,2)$ operating on $\mathcal{R}(s,n)/\mathcal{R}(k,n)$ is derived. Applying algorithm 1, we are able to compute the number of equivalent classes of cosets of Reed-Muller code $\mathcal{R}(1,n)$ when $n<11$, and the results are presented in Table \ref{37} and Table \ref{810}. Furthermore, we are able to compute the number of affine equivalent classes on $\mathcal{R}(s,n)/\mathcal{R}(k,n)$ when $-1\leq k< s\leq n\leq 10$, and the results are also presented in the paper. Section 4 concludes this paper.

\section{PRELIMINARIES}\label{section 2}
In this section, we provide some definitions and basic theories related to this paper. $\mathcal{R}(r,n)$ is the $r$-th order Reed-Muller code with length $2^n$, and we also use $\mathcal{R}(r,n)$ to denote the set of all Boolean functions with $n$ variables and degree at most $r$. 
\subsection{DEFINITIONS}
In addition to the truth table expression, every Boolean function can be uniquely represented as a polynomial, which is called the algebraic normal form (ANF) \cite{b15}.

\newtheorem{myDef}{Definition}
\newtheorem{theorem}{Theorem}
\newtheorem{remark}{Remark}

\begin{myDef}[algebraic normal form (ANF)]
	Every Boolean function $f:\mathcal{F}_2^n \rightarrow \mathcal{F}_2$ can be uniquely represented by a polynomial in n variables which has degree $\leq n$ 
	\begin{equation}  \label{ANF}
	f\left(x_{1}, \ldots, x_{n}\right)=\sum_{I \subseteq\{1, \ldots, n\}} a_{I} x^{I},
	\end{equation}
	where $x^{I}=\prod_{i \in I} x_{i}$,  $a_{I}$ takes value 0 or 1, $|I|$ is the degree of the monomial $ a_{I} x^{I}$. the addition and multiplication are calculated over $\mathcal{F}_2$.
\end{myDef}

Since every Boolean function is uniquely determined by its ANF,  we have a one-to-one correspondence between Boolean function and the coefficients $a_{I},I \subseteq\{1, \ldots, n\}$. By reordering the polynomial \eqref{ANF} in descending order of degree and variable indexes in ascending order, we can write the coefficients uniquely as a vector $(a_1,a_2,...,a_{2^n})^T$, which is called the coefficient vector of $f$. For instance, suppose the ANF of a Boolean function with three variables is $x_1x_2x_3\oplus x_1x_3\oplus x_2x_3\oplus x_3\oplus1$, then the coefficient vector is $(1,0,1,1,0,0,1,1)^T$. 

\begin{myDef}[coefficient vector space $\mathcal{CF}^n$ over $\mathcal{F}_2$]
	The coefficient vector of a Boolean function $f$, denoted as $cv(f)$, is a column vector with each element equaling to $a_I$ from \eqref{ANF} in descending order of degrees and ascending order of variable indexes. The coefficient vector space $\mathcal{CF}^n$ with dimension $2^n$ is a linear vector space constructed by all possible coefficient vectors of Boolean functions with $n$ variables.
\end{myDef}

\begin{myDef}[quotient coefficient vector space $\mathcal{CF}^n_{s,k}$ ]
	the coefficient vector space of $\mathcal{R}(s,n)$ is denoted by $\mathcal{CF}^n_s$, and the quotient coefficient vector space of $\mathcal{R}(s,n)/\mathcal{R}(k,n)$ is denoted by $\mathcal{CF}^n_{s,k}$. For simplicity, we define $\mathcal{R}(s,n)=\mathcal{R}(s,n)/\mathcal{R}(-1,n)$.
\end{myDef}
\par Let $f\in \mathcal{R}(s,n)$ be a Boolean function, then the order of monomials in ANF of $f$ must be less than or equal to $s$. In other words, the coefficients of monomials in ANF of $f$ whose orders are greater than $s$ must be zero, which reduces the dimension of $\mathcal{CF}^n_s$ from $2^n$ to $
\sum^s_{i=0}\left(\begin{IEEEeqnarraybox*}[][c]{,c,}
n\\
i
\end{IEEEeqnarraybox*}\right)$. Similarly, if $h\in \mathcal{R}(s,n)/\mathcal{R}(k,n)$ is a Boolean function, then the order of monomials in ANF of $f$ must be less than or equal to $s$ and greater than $k$, which means the dimension of $\mathcal{CF}^n_{s,k}$ is $
\sum^s_{i=k+1}\left(\begin{IEEEeqnarraybox*}[][c]{,c,}
n\\
i
\end{IEEEeqnarraybox*}\right)$.

\par  General linear group $\mathcal{GL}(n,2)$ is the set of all $n \times n$ invertible matrices over $\mathcal{F}_2$ and the affine group \cite{zhangyan} is defined to be: $$\mathcal{AGL}(n, 2)=\left\{(\mathbf{A}, \mathbf{b}) | \mathbf{A} \in \mathcal{GL}(n, 2), \mathbf{b} \in \mathcal{F}_{2}^{n}\right\}.$$ The size of $\mathcal{GL}(n,2)$ and $\mathcal{AGL}(n, 2)$ can be calculated as follows:
$$ |\mathcal{GL}(n, 2)|=\prod_{i=0}^{n-1}\left(2^{n}-2^{i}\right),$$
$$ |\mathcal{AGL}(n, 2)|=2^{n} \prod_{i=0}^{n-1}\left(2^{n}-2^{i}\right).$$
The affine group operates naturally on Boolean functions in the following way.

\begin{myDef}[equivalence on $\mathcal{R}(s,n)/\mathcal{R}(k,n)$]
	Two Boolean functions $f,h\in \mathcal{R}(s,n)$ are affine equivalent on $\mathcal{R}(s,n)/\mathcal{R}(k,n)$ if there exists an invertible matrix $A$ over $\mathcal{F}_2$, a vector $b\in \mathcal{F}_2^n$, and a Boolean function $r(x)\in \mathcal{R}(k,n)$ such that $h(x)=f(Ax\oplus b)\oplus r(x)$.
\end{myDef}
Affine equivalence of Boolean functions on $\mathcal{R}(s,n)$ can be viewed as a special case of affine equivalence of Boolean functions on $\mathcal{R}(s,n)/\mathcal{R}(-1,n)$. 
\begin{myDef}[$\mathcal{AGL}(n,2)$ operating on $\mathcal{R}(s,n)/\mathcal{R}(k,n)$]
	An element $g=(A,b) \in \mathcal{AGL}(n,2)$ operates on a Boolean function $f \in \mathcal{R}(s,n)/\mathcal{R}(k,n)$, transforming $f$ into $h$ satisfying $h(x)=f(Ax\oplus b)$ mod $\mathcal{R}(k,n)$.
\end{myDef}
\par In the above definition, $h\in \mathcal{R}(s,n)/\mathcal{R}(k,n)$, so $\mathcal{AGL}(n,2)$ operates on $\mathcal{R}(s,n)/\mathcal{R}(k,n)$. Computing the number of affine equivalent classes of Boolean function on $\mathcal{R}(s,n)/\mathcal{R}(k,n)$ is equal to computing the number of orbits of $\mathcal{AGL}(n,2)$ operating on $\mathcal{R}(s,n)/\mathcal{R}(k,n)$.

The work of \cite{zhangyan} calculated the number of affine equivalent classes of Boolean function on $\mathcal{R}(n,n)$, which is equivalently to computing the number of orbits of $\mathcal{AGL}(n,2)$ operating on the Boolean function set $\mathcal{R}(n,n)$. The author used Burnside’s lemma to calculate the number of orbits of a group $\mathcal{G}$ operating on a set $\mathcal{S}$. The number of orbits denoted by $N$ can be calculated as follows \cite{artin},
$$N=\frac{1}{|G|} \sum_{g \in G}|F i x(g)|,$$
where $Fix(g)=\{s \in \mathcal S | g s=s\}$.
\par To maintain consistency of this paper, we review some results in \cite{zhangyan} related to our method.  An integer $i(0\leq i\leq 2^n-1)$ can be expressed as $i=\sum_{k=1}^{n} b_{k} 2^{n-k}$ with $B(i)=(b_1,b_2,\cdots,b_n)^T$ being the vector expression of $i$ over $\mathcal{F}_2$. Also, an vector $v$ can be transformed to an integer $I(v)$ with $I(v)=\sum_{k=1}^{n} b_{k} 2^{n-k}$. Let $S=\{0,1,2,\cdots,2^{n}-1\}$, a mapping $\phi$ is defined from $\mathcal{AGL}(n,2)$ to a permutation group $P_n$ and the two groups are isomorphic:
\begin{equation*}
\begin{aligned} \phi: A G L(n, 2) & \rightarrow P_{n} \\ g & \mapsto \sigma_{g} \end{aligned},
\end{equation*} 
where $g=(A,b)$ and
$$\phi(g)=\sigma_{g}=\left[\begin{array}{cccc}{0} & {1} & {\cdots} & {2^{n}-1} \\ {\sigma_{g}(0)} & {\sigma_{g}(1)} & {\cdots} & {\sigma_{g}\left(2^{n}-1\right)}\end{array}\right]$$
$$\sigma_{g}(i)=I(g(B(i))), \quad i \in S.$$

\subsection{LINEAR REPRESENTATION OF $\mathcal{AGL}(n,2)$}
In this subsection, based on the theory of $\mathcal{AGL}(n,2)$ acting on $\mathcal{R}(s,n)/\mathcal{R}(k,n)$ \cite{b16}, \cite{xdh}, we describe the linear representations of $\mathcal{AGL}(n,2)$ on $\mathcal{CF}^n_{s,k}$ in our own definitions. A linear representation of a group $\mathcal{G}$ is a group operation on a vector space by invertible linear maps \cite{artin}. Consider the map $\rho_g$ defined as \begin{equation*}
\begin{aligned} \rho_g: \mathcal{CF}^n & \rightarrow \mathcal{CF}^n \\ cv(f) & \mapsto cv(g\circ f), \end{aligned}
\end{equation*} 
where $cv(f)$ is the coefficient vector of a Boolean function $f$ and $cv(g\circ f)$ is the coefficient vector of $g\circ f$.

The mapping $\rho_g$ defined above is an invertible linear operator on the vector space $\mathcal{CF}^n$ over $\mathcal{F}_2$.

The mapping $\rho:g \mapsto \rho_g$ is a group homomorphism: $\mathcal{AGL}(n,2) \mapsto \mathcal{GL}(\mathcal{CF}^n)$, where $\mathcal{GL}(\mathcal{CF}^n)$ is the group of invertible linear operators on $\mathcal{CF}^n$. Thus, $\rho$ is a representation of $\mathcal{AGL}(n,2)$ on the vector space $\mathcal{CF}^n$ over $\mathcal{F}_2$.

\par Although the vector space $\mathcal{CF}^n$ is chosen to construct the group representation, the vector space can be replaced by $\mathcal{CF}^n_{s,k}$ with $0\leq k< s\leq n$ to construct the corresponding representations of $\mathcal{AGL}(n,2)$. Consider the mapping $\tau_g$ defined as \begin{equation*}
\begin{aligned} \tau_g: \mathcal{CF}^n_{s,k} & \rightarrow \mathcal{CF}^n_{s,k} \\ cv_{s,k}(f) & \mapsto cv_{s,k}(g\circ f), \end{aligned}
\end{equation*} 
where $cv_{s,k}(f)$ is the coefficient vector of a Boolean function $f$ in $\mathcal{CF}^n_{s,k}$ and $cv_{s,k}(g\circ f)$ is the coefficient vector of $g\circ f$ in $\mathcal{CF}^n_{s,k}$.

The mapping $\tau:g \mapsto \tau_g$ is a group homomorphism: $\mathcal{AGL}(n,2) \mapsto \mathcal{GL}(\mathcal{CF}^n_{s,k})$, where $\mathcal{GL}(\mathcal{CF}^n_{s,k})$ is the group of invertible linear operators on $\mathcal{CF}^n_{s,k}$. Thus, $\tau$ is a representation of $\mathcal{AGL}(n,2)$ on the vector space $\mathcal{CF}^n_{s,k}$ over $\mathcal{F}_2$.

\par Now we are ready to construct the corresponding matrix representation of the invertible linear operator $\rho_g$ by choosing a specific basis of $\mathcal{CF}^n$. For computation simplicity, we choose the standard orthogonal basis $(e_1,e_2,...,e_{2^n})$, and $\rho_g$ can be expressed as a specific matrix. 
\begin{myDef}
	$\mathcal{M_{\rho}}$ is the set of all possible matrices constructed from all linear operators $\rho_g, g\in \mathcal{AGL}(n,2)$ by choosing the standard orthogonal basis.
\end{myDef} 
\par Clearly, $\mathcal{M_{\rho}}$ and $\mathcal{M_{\tau}}$ are matrix groups and both isomorphic to $\mathcal{AGL}(n,2)$. To show how the matrix is constructed, here is a specific example.
\newtheorem{example}{Example}
\begin{example}
	Let $A=\left(\begin{IEEEeqnarraybox*}[][c]{,c/c/c,}
	1 &1 &0\\
	0 &1 &0\\
	0 &0 &1
	\end{IEEEeqnarraybox*}\right)$, $b=\left(\begin{IEEEeqnarraybox*}[][c]{,c/c/c,}
	1 & 0 & 0
	\end{IEEEeqnarraybox*}\right)$, $g=(A,b)$
	and the coefficient vector $cv(f)=(c_1,c_2,c_3,c_4,c_5,c_6,c_7,c_8)^T$, so $f(x_1,x_2,x_3)=c_1x_1x_2x_3\oplus c_2x_1x_2\oplus c_3x_1x_3\oplus c_4x_2x_3\oplus c_5x_1\oplus c_6x_2\oplus c_7x_3\oplus c_8$. $\rho_g$ can be calculated in the following way. 
	First, we have $h(x)=f(Ax+b)=c_1x_1x_2x_3\oplus c_2x_1x_2\oplus c_3x_1x_3\oplus (c_3\oplus c_4)x_2x_3\oplus c_5x_1\oplus (c_5\oplus c_6)x_2\oplus (c_4\oplus c_7)x_3\oplus (c_5\oplus c_8)$. Then a matrix representation of $\rho_g$ can be easily obtained from \eqref{ex} by assigning the $8\times 8$ matrix to $\rho_g$ :
	\begin{equation} \label{ex}
	cv(h)=\left(\begin{IEEEeqnarraybox*}[][c]{,c/c/c/c/c/c/c/c,}
	1 &0 &0 &0 &0 &0 &0 &0\\
	0 &1 &0 &0 &0 &0 &0 &0\\
	0 &0 &1 &0 &0 &0 &0 &0\\
	0 &0 &1 &1 &0 &0 &0 &0\\
	0 &0 &0 &0 &1 &0 &0 &0\\
	0 &0 &0 &0 &1 &1 &0 &0\\
	0 &0 &0 &1 &0 &0 &1 &0\\
	0 &0 &0 &0 &1 &0 &0 &1
	\end{IEEEeqnarraybox*}\right)cv(f).
	\end{equation}
\end{example}
In fact, for any given $g=(A,b)$, we can calculate the matrix expression of $\rho_g$ in the following lemma. The lemma can be proved using linear algebra and the proof is omitted.
\newtheorem{lemma}{Lemma}
\begin{lemma}\label{basis}
	$\rho$ is the representation of $\mathcal{AGL}(n,2)$ on the vector space $\mathcal{CF}^n$ and $g\in \mathcal{AGL}(n,2)$. Let $\{e_1,e_2,...,e_{2^n}\}$ be the standard orthogonal basis of $\mathcal{CF}^n$ and $f_1,f_2,...,f_{2^n}$ be the corresponding Boolean functions of the basis. Then the $j$-th column of $\rho_g$ is $cv(g\circ f_i)$.
\end{lemma}
\par Similarly, we can calculate the matrix representation of $\mathcal{AGL}(n,2)$ on the vector space $\mathcal{CF}^n_{s,k}$ with Lemma 2.
\begin{lemma}\label{basis2}
	$\tau$ is the representation of $\mathcal{AGL}(n,2)$ on the vector space $\mathcal{CF}^n_{s,k}$ $g\in \mathcal{AGL}(n,2)$. Let $\{e_1,e_2,...,e_{d}\}$ be the standard orthogonal basis of $\mathcal{CF}^n_{s,k}$ whose dimension is $d$, and $f_1,f_2,...,f_{d}$ be the corresponding Boolean functions of the basis. Then the $j$-th column of $\tau_g$ is $cv_{s,k}(g\circ f_i)$.
\end{lemma}
\par In the next section, we use the symbol $\rho_g,\tau_g$ to denote the matrix expression of $\rho_g,\tau_g$ at the same time.

\section{COMPUTE THE NUMBER OF EQUIVALENT CLASSES ON $\mathcal{R}(s,n)/\mathcal{R}(k,n)$}
In this section, we present an efficient method to compute the equivalent classes on $\mathcal{R}(s,n)/\mathcal{R}(k,n)$. Since $\mathcal{R}(s,n)/\mathcal{R}(k,n)$ is isomorphic to $\mathcal{CF}^n_{s,k}$ and $\mathcal{AGL}(n,2)$ is isomorphic to $\mathcal{M_{\tau}}$, computing the number of equivalent classes on $\mathcal{R}(s,n)/\mathcal{R}(k,n)$ is equivalent to enumerate the orbits of $\mathcal{M_{\tau}}$ operating on $\mathcal{CF}^n_{s,k}$.

Consider the representation of the affine group $\mathcal{AGL}(n,2)$ on the vector space $\mathcal{CF}^n_{s,k}$, where $0\leq k< s\leq n$, we can take $\mathcal{CF}^n_{s,k}$ as a finite set and use Burnside's lemma to calculate the number of orbits of $\mathcal{M_{\tau}}$ operating on $\mathcal{CF}^n_{s,k}$. Computing the number of conjugate classes and representatives directly in $\mathcal{M_{\tau}}$ or $\mathcal{AGL}(n,2)$ can be source consuming, so we introduce an isomorphic permutation group $P_n$ described in section \ref{section 2} to reduce the complexity. The following lemma is used to demonstrate the isomorphism between $\mathcal{M_{\tau}}$ and $P_n$.

\begin{lemma}\label{property1}
	Let $\psi=\tau\phi^{-1}: P_n  \rightarrow \mathcal{M_{\tau}}$. then $\mathcal{M_{\tau}}$ and $P_n$ are isomorphic. The mappings between these two groups can be specified as follows. let $\sigma \in P_n$, then the corresponding element in $\mathcal{M_{\tau}}$ is $\tau_g$, where $g = \phi^{-1}(\sigma)$, and $g=(A,b) \in \mathcal{AGL}(n,2)$ can be calculated by taking $b=B(\sigma(0))$ and the $j$-th column of $A$ being $B(\sigma(2^{j-1}+1))\oplus b$.
\end{lemma}
\begin{IEEEproof}
	It is trivial to see that $\mathcal{M_{\tau}}$ and $P_n$ are isomorphic. The relationship between $\mathcal{AGL}(n,2)$, $\mathcal{M_{\tau}}$ and $P_n$ is demonstrated in Fig. \ref{f1}. It remains to check  $g = \phi^{-1}(\sigma)$. Since $B(0)$ is a zero vector and $\sigma(0)=I(\phi^{-1}(\sigma)(B(0)))=I(AB(0)\oplus b)=I(b)$, we have $b=B(\sigma(0))$. Let $\{e_i\}$ be the standard orthogonal basis of $\mathcal{F}^n_2$, we have $\sigma(I(e_j))=I(\phi^{-1}(\sigma)(e_j))=I(Ae_j\oplus b)$. Then $Ae_j=B(\sigma(2^{j-1}+1))\oplus b$.
\end{IEEEproof}

\begin{theorem}[counting formula]
	The number of equivalent classes on $\mathcal{R}(s,n)/\mathcal{R}(k,n)$ can be calculated as:
	\begin{equation}\label{counting}
	N_{s,k}=\frac{1}{|P_n|} \sum_{i=1}^c|\mathcal{CP}_i|\cdot2^{d-rank(\psi(\sigma_i)-I)},
	\end{equation}
	where $|P_n|=2^{n} \prod_{i=0}^{n-1}\left(2^{n}-2^{i}\right)$ is the size of the $P_n$, $c$ is the number of conjugate classes of $P_n$, $\mathcal{CP}_i$ is the $i$-th conjugate class of $P_n$, $\sigma_i$ is the representative of $\mathcal{CP}_i$, $d$ is the dimension of $\mathcal{CF}_{s,k}^n$, $I$ is the $d\times d$ identity matrix and $rank(\psi(\sigma_i)-I)$ is the rank of the matrix $\psi(\sigma_i)-I$.
\end{theorem}
\begin{IEEEproof}
	The way $g\in \mathcal{AGL}(n,2)$ operating on $\mathcal{R}(s,n)/\mathcal{R}(k,n)$ is the same as $\tau_g$ operating on $\mathcal{CF}^n_{s,k}$. Using Burnside's lemma, we have $$N_{s,k}=\frac{1}{|\mathcal{M_{\tau}}|} \sum_{\tau_g \in \mathcal{M_{\tau}}}|F i x(\tau_g)|.$$
	Since elements in the same conjugate class of a group have the same number of fixed elements in the set, by classifying $\mathcal{M_{\tau}}$ into different conjugate classes $\{\mathcal{CM}_1,\mathcal{CM}_2,...\mathcal{CM}_c\}$, $c$ is the number of conjugate classes of $\mathcal{M_{\tau}}$, we have $$N_{s,k}=\frac{1}{|\mathcal{M_{\tau}}|} \sum_{i=1}^c|\mathcal{CM}_i|\cdot|F i x(\tau_{g_i})|,$$
	where $\tau_{g_i}$ is a representative of $\mathcal{CM}_i$ and $|\mathcal{CM}_i|$ is the number of elements in $\mathcal{CM}_i$.
	Since isomorphic groups share the same number of conjugate classes and representatives of conjugate classes are in one-to-to correspondence, using Lemma \ref{property1}, we have $$N_{s,k}=\frac{1}{|P_n|} \sum_{i=1}^c|\mathcal{CP}_i|\cdot|F i x(\psi(\sigma_i))|,$$ where $\{\mathcal{CP}_1,\mathcal{CP}_2,...\mathcal{CP}_c\}$ is the conjugate class of $P_n$ and $\sigma_i$ is the corresponding representative. Since $|F i x(\psi(\sigma_i))|$ is equal to the number of solutions of $y\in \mathcal{CF}^n_{s,k}$ satisfying $\psi(\sigma_i)y=y$, we have $|F i x(\psi(\sigma_i))|=2^{d-rank(\psi(\sigma_i)-I)}$. This completes the proof.
\end{IEEEproof}
\begin{remark}
	When the affine group $\mathcal{AGL}(n,2)$ in Theorem 4 is replaced by its subgroups, the corresponding counting formula can be derived in the same way.
\end{remark}

\begin{remark}
	The complexity of computing the number of conjugate classes and representatives of $\mathcal{M_{\tau}}$ or $\mathcal{AGL}(n,2)$ can be so high that the computation process cannot be completed even when $n=8$. Instead of computing conjugate classes directly in $\mathcal{M_{\tau}}$ or $\mathcal{AGL}$, we introduce a permutation group $P_n$, which is isomorphic to $\mathcal{AGL}(n,2)$. As show in Fig. \ref{f1}, the three groups $\mathcal{AGL}(n,2)$, $\mathcal{M_{\tau}}$, and $P_n$ are isomorphic. In \cite{hulpke}, Hulpke proposed an algorithm to calculate conjugate classes of finite permutation groups by homomorphism mapping. The computation of conjugate classes in permutation groups is much easier than matrix groups, so we first compute the number of conjugate classes and representatives in $P_n$. Since isomorphic groups share the same number of conjugate classes and the same size of each class, the numbers $\mathcal{C}_i$ and $c$ in eq. \ref{counting} are obtained. The representatives of conjugate classes of $\mathcal{M_{\tau}}$ can be obtained by the representatives of $\mathcal{AGL}(n,2)$, and the representatives of $\mathcal{AGL}(n,2)$  can be obtained by the representatives of $P_n$ and Lemma \ref{property1}.
\end{remark}

\par Using the computation order described above and GAP \cite{gap}, we can calculate the number of equivalent classes up to 10 variables. The computation steps are summarized in Algorithm \ref{algorithm1}.

\begin{algorithm}
	\renewcommand{\algorithmicrequire}{\textbf{Input:}}
	\renewcommand{\algorithmicensure}{\textbf{Output:}}
	\caption{Computing the number of equivalent classes on $\mathcal{R}(s,n)/\mathcal{R}(k,n)$}
	\label{algorithm1}
	\label{alg:1}
	\begin{algorithmic}[1]
		\REQUIRE $n, s, k$
		\ENSURE the number of equivalent classes denoted as $N$
		\STATE let $N=0$, $d=dim(\mathcal{CF}^n_{s,k})$
		\STATE compute the conjugate classes of $P_n$ in GAP, find a representative $p_i$ in each conjugate class $\mathcal{C}_i$ and the number of elements $n_i$ in $\mathcal{C}_i$
		\FOR{each representative $p_i$}
		\STATE transform $p_i$ into $g_i=(A_i,b_i)$ using Property \ref{property1}
		\STATE transform $g_i$ into $\tau_{g_i}$ using Lemma \ref{basis2}
		\STATE compute the rank of $\tau_{g_i}-I$, denoted as $r_i$
		\STATE $N+=n_i\times 2^{d-r_i}$
		\ENDFOR
		\STATE $N=N/|\mathcal{M_{\tau}}|$
		\STATE return $N$ as the number of equivalent classes on $\mathcal{R}(s,n)/\mathcal{R}(k,n)$
	\end{algorithmic}  
\end{algorithm}

\par Consider the vector space $\mathcal{CF}^n_s$. When $s=n$, we can use Algorithm \ref{algorithm1} to compute the number of affine equivalent classes of Boolean functions, which has been done in \cite{zhangyan} using a different approach. Thus, the experiment results are omitted here in this paper. When $s<n$, we can use Algorithm \ref{algorithm1} to compute the number of affine equivalent classes of Boolean functions whose degree are less than or equal to $s$.
\par Now consider the vector space $\mathcal{CF}^n_{s,k}$, 
\begin{enumerate}[\IEEEsetlabelwidth{12)}]
	\item when $s=n, k=0$, we classify Boolean functions under this equivalence definition that $f,h\in \mathcal{BS}^n$ are affine equivalent if there exists $A\in \mathcal{GL}(n,2)$, $b\in \mathcal{F}^n_2$, $d\in \mathcal{F}_2$ such that $h(x)=f(Ax+b)+d$. The affine equivalence of Boolean functions under this definition can be calculated using method in \cite{zhangju} with NP group replaced by affine group.
	\item when $s=n, k=1$, we use Algorithm \ref{algorithm1} to compute the number of equivalent classes of Boolean functions modulo linear functions.  Classifying Boolean functions modulo linear functions is equivalent to computing the classification of cosets of Reed-Muller code $R(1,n)$.
	\item when $s,k$ take other values, we use Algorithm \ref{algorithm1} to compute the number of equivalent classes of Boolean functions on $\mathcal{R}(s,n)/\mathcal{R}(k,n)$.
\end{enumerate}
\subsection{Compute the Number of Classes of the Cosets of $\mathcal{R}(1,n)$ }
Based on previous publications, the classification of the cosets of $R(1,n)$ has been calculated when $n\leq 7$. Using the counting formula \eqref{counting}, we are able to compute the number of classes up to 10 variables. Classifying Boolean functions modulo linear functions is equivalent to computing the classification of cosets of Reed-Muller code $R(1,n)$, which is equivalent to enumerating orbits of $\mathcal{AGL}(n,2)$ operating on $\mathcal{CF}^n_{n,1}$. The results are presented in Table \ref{37} and Table \ref{810}.

\begin{table}  
	\renewcommand\arraystretch{1.8}
	\caption{the number of classes of the cosets of $R(1,n)$ for $n\leq 7$}  
	\label{37}
	\centering
	\begin{tabular*}{8.5cm}{cccccc}  
		\hline  
		n & 3  & 4 &5 &6 & 7 \\  
		\hline  
		classes & 3 & 8 &48 &150357 &63379147320777408548\\  
		\hline  
	\end{tabular*}  
\end{table}

\begin{table}  
	\renewcommand\arraystretch{1.8}
	\caption{the number of classes of the cosets of $R(1,n)$ for $8\leq n\leq 10$}  
	\label{810}
	\centering
	\begin{tabular*}{8.9cm}{cccccc}  
		\hline  
		n & classes \\  
		\hline  
		8 &165185676632164460815766870775791408749821298442607120 \\  
		\hline  
		\multirow{3}*{9} & 365536222811389843514666387723835414818483436058712103\\&049956307039407351597779452899567037213641051913718290\\&46662966131452802\\
		\hline
		\multirow{5}*{10}&233928218689469200516884786693734036891557193749847043\\&210506069040915217122543732791617959958627227131213293\\&821649344237210406866167441313526703679920986684370312\\&072776334853708791163806695794267614394278266564198718\\&824702675531267798195533299157850647374756378184100860950\\
		\hline
	\end{tabular*}  
\end{table}

\subsection{Compute the Number of Affine Equivalent Classes of Boolean Functions on $\mathcal{R}(s,n)/\mathcal{R}(k,n)$}
We have computed the number of equivalent classes on $\mathcal{R}(s,n)/\mathcal{R}(k,n)$ when $-1\leq k<s\leq n\leq 10$. The experiment results are presented in Tables \ref{n7}-\ref{n103}. There exists a symmetry that the number of equivalent classes on $\mathcal{R}(s,n)/\mathcal{R}(k,n)$ is equal to the number of equivalent classes on $\mathcal{R}(n-1-k,n)/\mathcal{R}(n-1-s,n)$, which has been proved in \cite{xdh}. In Tables \ref{n7}-\ref{n103}, we only show half of the results, while the rest can be obtained by symmetry.

\section{Conclusion}
We present a method to compute the number of equivalent classes on $\mathcal{R}(s,n)/\mathcal{R}(k,n)$ when $n<11$. An even more challenging problem is to find representatives in each class. The method used in the paper may be applied to find the representative Boolean function in each class and helps to detect affine equivalence of Boolean functions, which are the highlights of some future work to focus on.

\begin{table}  
	\renewcommand\arraystretch{1.7}
	\caption {the number of equivalent classes on $\mathcal{R}(s,n)/\mathcal{R}(k,n)$ when $n=7$} 
	\label{n7}
	\centering
	\begin{tabular*}{8.8cm}{p{0.3cm}p{8.5cm}<{\centering}}   
		\hline  
		(k,s) & classes \\  
		\hline  
		(0,1) &2 \\  
		\hline  
		(0,2) &8 \\
		\hline
		(0,3) &1890\\
		\hline
		(0,4) &15115039412866\\
		\hline
		(0,5) &31689573670826669852\\
		\hline
		(0,6) &4056249792080063701952\\
		\hline
		(0,7) &8112499583888855378066\\
		\hline
		(1,2) &4\\
		\hline
		(1,3)&179\\
		\hline
		(1,4)&118140881980\\
		\hline
		(1,5)&247576791326613880\\
		\hline
		(1,6)&31689573670826669852\\
		\hline
		(1,7)&63379147320777408548\\
		\hline
		(2,3)&12\\
		\hline
		(2,4)&68443\\
		\hline
		(2,5)&118140881980\\
		\hline
		(2,6)&15115039412866\\
		\hline
		(2,7)&30230045341814\\
		\hline
		(3,4)&12\\
		\hline
		(3,5)&179\\
		\hline
		(3,6)&1890\\
		\hline
		(3,7)&3486\\
		\hline
		(4,7)&12\\
		\hline
		(5,7)&3\\
		\hline
		(6,7)&2\\
		\hline
	\end{tabular*}  
\end{table}

\begin{table}  [H]
	\renewcommand\arraystretch{1.5}
	\caption {the number of equivalent classes on $\mathcal{R}(s,n)/\mathcal{R}(k,n)$ when $n=8$} 
	\label{n8}
	\begin{tabular*}{9.1cm}{p{0.3cm} c }    
		\hline  
		(k,s) & classes \\  
		\hline  
		(0,1) &2 \\  
		\hline  
		(0,2) &9 \\
		\hline
		(0,3) &3814830\\
		\hline
		(0,4) &4269949724986494593784770116\\
		\hline
		(0,5) &307682299301470708167866113375397397427671066\\
		\hline
		(0,6) &82592838316082230407883435898918419677493521833649626\\
		\hline
		(0,7) &21143766608916976744527389019561070737773058189004193090\\
		\hline
		(0,8) &42287533217833953489054778023401252726576585396037133766\\
		\hline
		(1,2) &5\\
		\hline
		(1,3)&20748\\
		\hline
		(1,4)&16679491361446456200861682\\
		\hline
		(1,5)&1201883981646910151672305888222106927574728\\
		\hline
		(1,6)&322628274672198478156638301650788590555856626021206\\
		\hline
		(1,7)&82592838316082230407883435898918419677493521833649626\\
		\hline
		(1,8)&165185676632164460815766870775791408749821298442607120\\
		\hline
		(2,3)&32\\
		\hline
		(2,4)&62136089224841664\\
		\hline
		(2,5)&4477366736959738376608731157309430\\
		\hline
		(2,6)&1201883981646910151672305888222106927574728\\
		\hline
		(2,7)&307682299301470708167866113375397397427671066\\
		\hline
		(2,8)&615364598602941416335607372199130813176024448\\
		\hline
		(3,4)&999\\
		\hline
		(3,5)&62136089224841664\\
		\hline
		(3,6)&16679491361446456200861682\\
		\hline
		(3,7)&4269949724986494593784770116\\
		\hline
		(3,8)&8539899449972486788122917594\\
		\hline
		(4,8)&7611801\\
		\hline
		(5,8)&14\\
		\hline
		(6,8)&3\\
		\hline
		(7,8)&2\\
		\hline
	
	\end{tabular*}  
\end{table}

\section*{Acknowledgment}

This work was supported by the National Natural Science Foundation of China under Grant No. 61572109.

\ifCLASSOPTIONcaptionsoff
  \newpage
\fi

\begin{IEEEbiographynophoto}{Xiao Zeng}
	received the B.S. degree in mathematics and applied mathematics from University of Electronic Science and Technology of China, China. He is currently pursuing the Ph.D. with the School of Computer Science and Engineering, University of Electronic Science and Technology of China, China. His research interests include Boolean functions, machine learning and quantum computing.
\end{IEEEbiographynophoto}

\begin{IEEEbiographynophoto}{Guowu Yang} received the B.S. degree from the University of Science and Technology of China in 1989, the M.S. degree from the Wuhan University of Technology in 1994, and the Ph.D. degree in electrical and computer engineering from Portland State University in 2005. He was with the Wuhan University of Technology from 1989 to 2001 and with Portland State University from 2005 to 2006. He is currently a Full Professor with the University of Electronic Science and Technology of China. His research interests include verification, logic synthesis, quantum computing, and machine learning. He has published over 100 journal and conference papers.
\end{IEEEbiographynophoto}


\begin{table}  [H]
	\renewcommand\arraystretch{1.7}
	\caption{the number of equivalent classes on $\mathcal{R}(s,n)/\mathcal{R}(k,n)$ when $n=9$} 
	\label{n9}
	\begin{tabular*}{9.1cm}{p{0.3cm} c }   
		\hline  
		(k,s) & classes \\  
		\hline  
		(0,1) &2 \\  
		\hline  
		(0,2) &10\\
		\hline
		(0,3) &1901093335846\\
		\hline
		(0,4) &161629820579418429077416198759587963832041903792286\\
		\hline
		\multirow{2}*{(0,5)} &137499444779425838856562316135234629055974069079270\\&69890926662777447694197527953076865740\\
		\hline
		\multirow{3}*{(0,6)} &2659626063624381229708279626530613476476394655583033\\&8745748187394431227885639909041560101790620430691219470\\&1328830\\
		\hline
		\multirow{3}*{(0,7)} &1827681114056949217573331938619177074092417180293560\\&5152497815625351827227338191438289523993699895699112563\\&373313191508527234\\	
		\hline
		\multirow{3}*{(0,8)} &9357727303971579993975459525730179452607115169857023\\&0760700206553067575034105725408366957187100957673358370\\&10888560978005336426\\
		\hline
		\multirow{3}*{(0,9)} &
		18715454607943159987950919051460358905214230339714046\\&15214004129654400502381741403609041837822670777886838609\\&3987560627143587626\\
		\hline
		(1,2) &5\\
		\hline
		(1,3) &3718776534\\
		\hline
		(1,4) &315683243319181079202852900667960591620204925846\\	
		\hline		
		\multirow{2}*{(1,5)} &268553603084816091516723384122836898396394316372\\&17153057359086171274280119527124951022\\	
		\hline
		\multirow{2}*{(1,6)} &519458215551636958927398364556862290211346521386464042551\\&640156788983185152596618356881991077880623268520470062\\	
		\hline
		\multirow{3}*{(1,7)} &356968967589247894057291394261558569063007261261745877724\\&7936060038457029184656577443450661944396194853024347309\\&2192802568\\	
		\hline
		\multirow{3}*{(1,8)} &182768111405694921757333193861917707409241718029356051524\\&9781562535182722733819143828952399369989569911256337331\\&3191508527234\\	
		\hline
		\multirow{3}*{(1,9)} & 365536222811389843514666387723835414818483436058712103\\&049956307039407351597779452899567037213641051913718290\\&46662966131452802\\
		\hline
		(2,3) &349\\	
		\hline
		(2,4) &4593795795813149805203372253161806928\\	
		\hline
		\multirow{2}*{(2,5)} &3907969266362722432193570873779909768911853476867\\&17279694254327466286185670\\	
		\hline
		\multirow{2}*{(2,6)} &75591119173861364677926520319703714532411062241\\&97786158143800438207494311103016415060904989004678658\\	
		\hline
		\multirow{3}*{(2,7)} &51945821555163695892739836455686229021134652138\\&64640425516401567889831851525966183568819910778806232\\&68520470062\\	
		\hline
	\end{tabular*}  
\end{table}  

\begin{table}  
	\renewcommand\arraystretch{1.7}
	\caption{continued Table of $n=9$}  
	\label{n91}
	\begin{tabular*}{9.1cm}{p{0.3cm} c }   
		\hline  
		(k,s) & classes \\  
		\hline
		\multirow{2}*{(2,8)} &265962606362438122970827962653061347647639465558303387457\\&481873944312278856399090415601017906204306912194701328830\\	
		\hline
		\multirow{2}*{ (2,9)} &5319252127248762459416559253061226952952789311166067749082\\&14706386097145644009393712741058554607546494018455194978\\
		\hline	
		(3,4) &121597673132830\\	
		\hline
		(3,5) &20203727572427022670501203228923012633782403068610\\	
		\hline
		\multirow{2}*{(3,6)} 
		&390796926636272243219357087377990976891185347686717279694\\&254327466286185670\\	
		\hline
		\multirow{2}*{(3,7)} 
		&268553603084816091516723384122836898396394316372171530573\\&59086171274280119527124951022\\	
		\hline
		\multirow{2}*{(3,8)} 
		&137499444779425838856562316135234629055974069079270698909\\&26662777447694197527953076865740\\	
		\hline
		\multirow{2}*{(3,9)} 
		&274998889558851677713124632270469258111948127490148104446\\&23477988535234386333472612284760\\	
		\hline
		(4,9) &323259641158836858154832039212134948029474970127110\\
		\hline	
		(5,9) &3802173825434\\	
		\hline
		(6,9) &15\\	
		\hline
		(7,9) &3\\	
		\hline
		(8,9) &2\\	
		\hline
	\end{tabular*}  
\end{table}

\begin{table}  
	\renewcommand\arraystretch{1.7}
	\caption{the number of equivalent classes on $\mathcal{R}(s,n)/\mathcal{R}(k,n)$ when $n=10$} 
	\label{n100}
	\begin{tabular*}{9.1cm}{p{0.3cm} c }   
		\hline  
		(k,s) & classes \\  
		\hline
		(0,1) & 2\\
		\hline
		(0,2) & 11\\
		\hline
		(0,3) & 127629062074904087575\\
		\hline
		\multirow{2}*{(0,4)}& 
		2100125928327007558293266973607225059836641678232597040\\&03327277105602635817744207388\\
		\hline
		\multirow{3}*{(0,5)}& 
		1519862305640264846965634643730038918627098653703425490\\&9635762962265114529220598858983717861217764025565681619\\&47545787344521281779265482179345504677511354183306\\
		\hline
		\multirow{4}*{(0,6)}&
		25009403504317715641840949950852195854692951488570311964\\&093085539730925721254067453954334236393194076417293468\\&07020642479015694946040534496307919492759143351136590397\\&350154346444734136580473767877090380759849150536753040226\\
		\hline
	\end{tabular*}  
\end{table}

\begin{table}  
	\renewcommand\arraystretch{1.7}
	\caption{continued table of $n=10$}  
	\label{n101}
	\begin{tabular*}{9.1cm}{p{0.3cm} c }   
		\hline  
		(k,s) & classes \\  
		\hline
		\multirow{5}*{(0,7)} & 33243199295820488787979455010732584254861477823773021101\\&8438906536568274436654021374580379139260458830183316390\\&276693696407270108641293883895142679091700503064268337\\&9753206807582317650902121486047431495975858238390507479\\&667107179054450175682041572072189950284\\
		\hline
		\multirow{5}*{(0,8)} & 11696410934473460025844239334686701844577859687492352160\\&5253034520457608561271866395808979979313613565606646910\\&8246721186052034330837206773598425531605665212473768799\\&473801842811676072161980197326943032073534424834295653\\&38771546774916501908837963411126439523824201438012706\\
		\hline
		\multirow{5}*{(0,9)} & 11977124796900823066464501078719182688847728319992168612\\&3779107348948590085307906295923307209038148250107717683\\&4934118914685814883824850954006801662217858770958819577\\&974347837408341182778404307853297753469370235501412508\\&32947381490374527840181122222654290776710468299239327504\\
		\hline
		\multirow{5}*{(0,10)} &
		23954249593801646132929002157438365377695456639984337224\\&7558214697897180170615812591846614418076296500215435366\\&9868237829371629767649701895374967903398328398547114544\\&514988923628570755092216384254539092723383821526404189\\&57275045327945473388097446782728753447682887217701463548\\
		\hline
		(1,2) & 6\\
		\hline
		(1,3) & 124638757107379571\\
		\hline
		\multirow{2}*{(1,4)} & 
		205090422688184331864598504108343243638438903051037979779\\&728144928709583019921334\\
		\hline
		\multirow{3}*{(1,5)} & 
		148424053285182113961487758176761613147177616172882552946\\&236051117700844671810307072624776875008139607818145662\\&1171566506560404984305403663936024703356725374\\
		\hline
		\multirow{4}*{(1,6)} & 
		244232456096852691814853026863790975143485854380569452774\\&346841556431411296643704470688021645802782792299601402\\&48167380566364507938514357931236358006951284646833602004\\&89328753033725085653262184824544170756512078915566180\\
		\hline
		\multirow{5}*{(1,7)} & 
		324640618123246960820111865339185393113881619372783409197\\&694244664618956296375883101303136935941897732706829641\\&80545643010531119193838171523741847060767256840811691083\\&54092170350735530250729637560325726303464828903224861\\&167941027505802054004991449154124838\\
		\hline
		\multirow{5}*{(1,8)} & 
		1142227630319673830648851497527998227009556610106675015676\\&29916523884385423289643264845439985633015034989363968\\&2229326819945965401385013424844388519570333660747073800836\\&775809014476432590169881459969687869329163981347733\\&64668659167318560642093291868632343311726513714396\\
		\hline
	\end{tabular*}  
\end{table} 

\begin{table}  
	\renewcommand\arraystretch{1.7}
	\caption{continued table of $n=10$}  
	\label{n102}
	\begin{tabular*}{9.1cm}{p{0.3cm} c }   
		\hline  
		(k,s) & classes \\  
		\hline
		\multirow{5}*{(1,9)} & 116964109344734600258442393346867018445778596874923521605\\&253034520457608561271866395808979979313613565606646910824\\&6721186052034330837206773598425531605665212473768799473801\\&842811676072161980197326943032073534424834295653387715467\\&74916501908837963411126439523824201438012706\\
		\hline
		\multirow{5}*{(1,10)}&
		233928218689469200516884786693734036891557193749847043210\\&506069040915217122543732791617959958627227131213293821649\\&3442372104068661674413135267036799209866843703120727763348\\&537087911638066957942676143942782665641987188247026755312\\&67798195533299157850647374756378184100860950\\
		\hline
		(2,3) & 3691561\\
		\hline
		\multirow{2}*{(2,4)} & 
		582902040060231263042997715290257412831729732030215811242\\&5987940194\\
		\hline
		\multirow{3}*{(2,5)} & 
		421846531495424738868359094615134418822846058029541321024\\&258814432173839728599928080518081750147253114085099139540\\&30914479041928106993313180762\\
		\hline
		\multirow{4}*{(2,6)} & 
		694150390065865087236866846932886617035722606432754710040\\&459865552100522188081924157932921330001483764869771203591\\&963717420912059825357017483021633380913683326254747713665\\&41395818249961735009311931925187892\\
		\hline
		\multirow{5}*{(2,7)} &
		922684131760567427193427264851338216868389336505013628759\\&838424757004873214769597134763636297662884905840498729076\\&291511122422062996761719795360542451418474583483372483532\\&8967858927424546310274748589879568986141296930044350358750\\&0892897327426\\
		\hline
		\multirow{5}*{(2,8)} &
		324640618123246960820111865339185393113881619372783409197\\&694244664618956296375883101303136935941897732706829641805\\&456430105311191938381715237418470607672568408116910835409\\&2170350735530250729637560325726303464828903224861167941027\\&505802054004991449154124838\\
		\hline
		\multirow{5}*{(2,9)} &
		332431992958204887879794550107325842548614778237730211018\\&438906536568274436654021374580379139260458830183316390276\\&693696407270108641293883895142679091700503064268337975320\\&6807582317650902121486047431495975858238390507479667107179\\&054450175682041572072189950284\\
		\hline
		\multirow{5}*{(2,10)} &
		664863985916409775759589100214651685097229556475460422036\\&877813073136548873308042749160758278520917660366632780553\\&387392814540215316528976598881150209781432529192229667522\\&7265422303736049071478675158728985365906879676023485563367\\&382084859367566028202211363624\\
		\hline
	\end{tabular*}  
\end{table} 

\begin{table}  
	\renewcommand\arraystretch{1.7}
	\caption{continued table of $n=10$}  
	\label{n103}
	\begin{tabular*}{9.1cm}{p{0.3cm} c }   
		\hline  
		(k,s) & classes \\  
		\hline
		(3,4) & 4490513974418226922710218421015600\\
		\hline
		\multirow{2}*{(3,5)} &
		317362057399582697291910046772156029664403537781330998408\\&73223280366018289714092671398201946992247091347326\\
		\hline
		\multirow{3}*{(3,6)} &
		522220711771847508679667650456410288664847468264583945134\\&495391511096729189539298872044188900478664171424947043174\\&53092880619313017642491652572876678931531251584335846086\\
		\hline
		\multirow{4}*{(3,7)} &
		694150390065865087236866846932886617035722606432754710040\\&459865552100522188081924157932921330001483764869771203591\\&963717420912059825357017483021633380913683326254747713665\\&41395818249961735009311931925187892\\
		\hline
		\multirow{4}*{(3,8)} &
		244232456096852691814853026863790975143485854380569452774\\&346841556431411296643704470688021645802782792299601402481\\&673805663645079385143579312363580069512846468336020048932\\&8753033725085653262184824544170756512078915566180\\
		\hline
		\multirow{4}*{(3,9)} &
		250094035043177156418409499508521958546929514885703119640\\&930855397309257212540674539543342363931940764172934680702\\&064247901569494604053449630791949275914335113659039735015\\&4346444734136580473767877090380759849150536753040226\\
		\hline
		\multirow{4}*{(3,10)} &
		500188070086354312836818999017043917093859029771406239281\\&861710794618514425081349079086684727863881528345869359190\\&112116345569416509939451172509533790412157911629068950763\\&7687889202045551156761584269684153298216997045167494\\
		\hline
		(4,5) & 19749489318110485970697971583208968127316501515\\
		\hline
		\multirow{2}*{(4,6)} &
		317362057399582697291910046772156029664403537781330998408\\&73223280366018289714092671398201946992247091347326\\
		\hline
		\multirow{3}*{(4,7)} &
		421846531495424738868359094615134418822846058029541321024\\&258814432173839728599928080518081750147253114085099139540\\&30914479041928106993313180762\\
		\hline
		\multirow{3}*{(4,8)} &
		148424053285182113961487758176761613147177616172882552946\\&236051117700844671810307072624776875008139607818145662117\\&1566506560404984305403663936024703356725374\\
		\hline
		\multirow{3}*{(4,9)} &
		151986230564026484696563464373003891862709865370342549096\\&357629622651145292205988589837178612177640255656816194754\\&5787344521281779265482179345504677511354183306\\
		\hline
		\multirow{3}*{(4,10)} &
		303972461128052969393126928746007783725419730740685098192\\&715259245302290581726309526116983204075601826300635522671\\&6023176032557448172538058658463501132247449478\\
		\hline
		\multirow{3}*{(5,10)} &
		420025185665401511658653394721445011651629620152235205656\\&901741950144699908180945624\\
		\hline
		(6,10) & 255258123991385004386\\
		\hline
		(7,10) & 17\\
		\hline
		(8,10) & 3\\
		\hline
		(9,10) & 2\\
		\hline
	\end{tabular*}  
\end{table} 


\end{document}